# LLM-Guided Open Hypothesis Learning from Autonomous Scanning Probe Microscopy Experiments

Boris Slautin[*], Utkarsh Pratiush, Yu Liu, Kamyar Barakati, Sergei Kalinin[*]

Department of Materials Science and Engineering, University of Tennessee, Knoxville, TN 37923, USA

## Abstract

Autonomous experimentation has transformed microscopy and materials discovery by enabling closed-loop optimization including imaging and spectroscopy tuning, strucutre property relationship discovery, and exploration of combinatorial libraries. However, most current workflows remain limited to selecting measurements within fixed objective or hypothesis spaces, rather than generating new physical models from experimental data. Here, we introduce an open hypothesis-learning framework that combines symbolic regression with large-language-model-based physical evaluation and implement it for autonomous scanning probe microscopy. Symbolic regression generates candidate analytical relationships directly from sparse measurements, while the language-model evaluator ranks these candidates according to physical plausibility, scaling behavior, and consistency with known mechanisms. We demonstrate the approach on autonomous piezoresponse force microscopy measurements of ferroelectric domain switching in a PZT thin film. Starting from five seed measurements, the workflow evolves from physically incomplete candidate expressions toward interpretable voltage–time growth laws consistent with kinetic domain-wall motion. This work extends autonomous microscopy from closed-loop optimization toward open hypothesis discovery, where candidate physical laws emerge from the experiment itself rather than being specified in advance. More broadly, the framework establishes a route for integrating symbolic regression, physical reasoning, and adaptive experimentation into hierarchical autonomous scientific workflows.

[*] Authors to whom correspondence should be addressed: bslauti1@utk.edu and sergei2@utk.edu

## I. Introduction

Autonomous experimentation has rapidly evolved into a well-established paradigm across microscopy, materials science, physics, and chemistry, with substantial progress over the past decade in closing experimental loops through machine learning–driven optimization.[1-10] In particular, Bayesian optimization (BO) has become the standard for closed-loop experimental design across diverse domains, including a wide range of microscopy applications, materials synthesis, and catalysis, where it enables data-efficient exploration of complex parameter spaces, often relying on additional structure or embeddings to remain effective in higher dimensions.[6, 11-18] For microscopy, these include the optimization of imaging and spectroscopy, structure-property relationship discovery, and materials discovery in combinatorial spread and random libraries. However, current implementations remain largely restricted to closed optimization paradigms, where predefined objectives guide exploration and physical understanding is imposed as prior knowledge rather than learned from data. Consequently, while such systems efficiently converge to optima within predefined search spaces such as instrument parameters, microstructures, or composition, they lack mechanisms for open hypothesis generation and physically grounded interpretation (Figure 1a).

The next natural step in the autonomous experimentation lies in the transition toward more open and hierarchical forms of autonomous experimentation.[19-23] In such settings, experimental workflows extend beyond predefined objectives to include sequential decision-making, knowledge transfer, and coordination across multiple levels of abstraction. Recent advances in large language models and agent-based systems provide a pathway toward such architectures, enabling the integration of reasoning, planning, and experimental control within autonomous loops.[21, 23-28] Emerging demonstrations of LLM-driven laboratory assistants and multi-agent systems highlight the potential for these approaches to orchestrate complex scientific workflows and to transform experimental platforms from passive measurement tools into active participants in the discovery process.[23, 29, 30] Importantly, such hierarchical workflows critically rely on the ability to extract and represent interpretable knowledge from data. Without this capability, information cannot be effectively propagated across decision layers or between agents. This makes open hypothesis formulation where hypotheses are generated, evaluated, and refined as outcomes of the experimental process particularly crucial for emerging autonomous systems.

Hypothesis-learning components have recently been incorporated into autonomous microscopy frameworks to accelerate optimization and enable model-based interpretation of experimental outcomes.[11, 31, 32] In these implementations, the experimental loop is not driven only by interpolation or maximization of a measured response, but by Bayesian updating over a set of competing physical models embedded into the workflow. This strategy has been demonstrated in autonomous SPM studies of ferroelectric domain switching, where predefined domain-growth hypotheses were evaluated during the experiment guided by the classical BO loop. The hypothesis space (candidates) was specified prior to the experiment and then explored within a single instrument-sample pair, enabling rapid iteration on experimentally accessible timescales. This represents an important step beyond purely data-driven optimization. However, the formulation of the candidate hypotheses remains external to the autonomous loop: the system selects among human-defined models rather than generating new physical relationships directly from data. This creates a fundamental gap for hierarchical autonomous experimentation, where interpretable knowledge must be produced, evaluated, and propagated across decision layers.[31] Here, we define open hypothesis learning as the ability to generate, evaluate, and refine candidate physical relationships directly from experimental data and prior knowledge, without relying on a predefined set of human-designed model classes.

Here, we introduce an approach for open hypothesis learning in autonomous microscopy, where candidate models are generated directly from experimental data using symbolic regression and subsequently evaluated through structured LLM-based reasoning to identify physically consistent and interpretable relationships. We note that the relative roles of LLMs and verification can be inverted, giving rise to multiple scenarios of hypothesis experiment planning.[21, 33, 34] Unlike prior approaches that rely on predefined hypothesis spaces, the proposed framework enables hypothesis generation and selection as an intrinsic outcome of the experimental process. The approach is demonstrated on autonomous scanning probe microscopy experiments probing ferroelectric switching dynamics. This establishes a pathway toward integrating data-driven modeling with physics-aware reasoning in autonomous experimental workflows.

## II. Hypothesis Learning Framework

The proposed open hypothesis learning framework is based on two core components: (i) a symbolic regression module that infers analytical expressions directly from experimental data, and (ii) an LLM-based evaluator that assesses the physical consistency of candidate models. Together, these components enable generation and selection of interpretable hypotheses as an intrinsic part of the experimental process (Figure 1b).

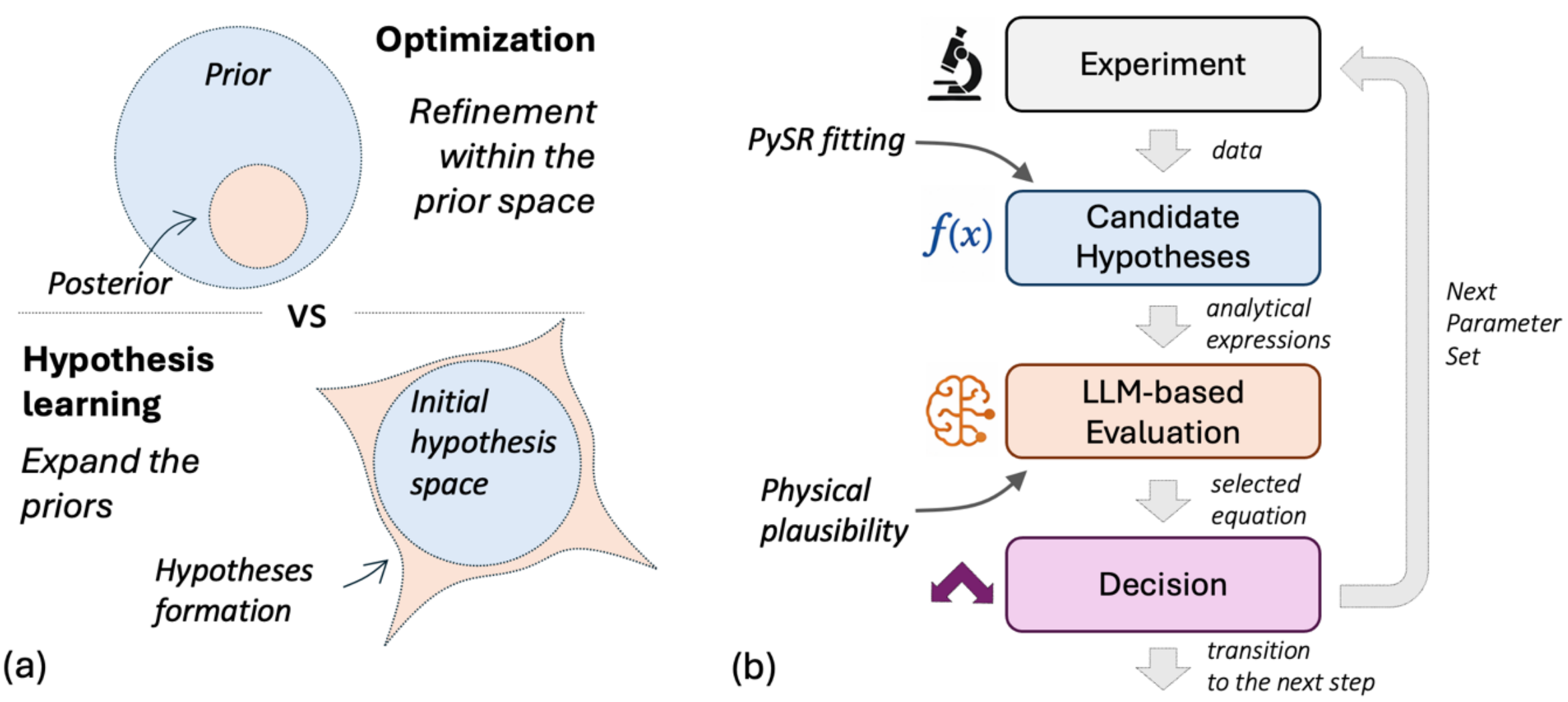

**Figure 1.** (a) Schematic illustration of hypothesis-space expansion. In conventional Bayesian optimization, measurements refine predictions within a fixed search space. In hypothesis learning, symbolic regression and LLM-based evaluation are used to expand the set of plausible functional forms considered during the experiment. (b) Schematic of the open hypothesis learning module.

The workflow proceeds as follows (Figure 1b). First, experimental data are acquired within an autonomous measurement loop. These data are used as input to a symbolic regression procedure, which constructs a set of candidate analytical expressions representing possible relationships between the measured variables. This results in a hypothesis space comprising multiple models with comparable predictive performance but distinct functional forms.

In the next step, the candidate models are evaluated using a structured LLM-based reasoning module. The LLM is provided with the candidate equations, variable definitions, and relevant physical context, and is tasked with assessing their consistency with known physical principles. This evaluation produces a ranking of hypotheses, accompanied by scores and reasoning that reflect physical plausibility, expected scaling behavior, and extrapolation stability.

Based on this evaluation, a physically consistent and interpretable hypothesis is selected. This hypothesis, together with the associated reasoning, forms the basis for the subsequent decision-making step. Specifically, the selected model can be used either to guide acquisition of additional data for model refinement, for example through optimization of residual uncertainty, or to inform transitions within a broader experimental workflow where the inferred relationship is used as a compact representation of the observed behavior. In this step, it is assumed that all physics that can be learned with the given combination of the sample and instrument has been learned, and the experimental planning should incorporate new samples or measurement modalities, or incorporated into fundamental knowledge base.

### II.1 Symbolic Regression for Hypothesis Generation

The search for compact analytical relationships has historically been central to physical science, where empirical observations are often transformed into mathematical laws. Symbolic regression offers a computational analog of this process by searching directly over the space of mathematical expressions rather than fitting parameters within a predefined model form. This process is necessarily incomplete, since the accessible hypotheses are constrained by data quality, operator choice, noise, and imposed complexity limits. Nevertheless, symbolic regression provides a practical mechanism for open hypothesis generation: it produces explicit candidate equations that can be interpreted, ranked, and tested against physical expectations. In contrast to conventional regression methods, which assume a fixed model structure, symbolic regression simultaneously identifies both the functional form of the relationship and its associated parameters.

In practice, symbolic regression constructs a population of candidate expressions and iteratively evolves them using operations such as mutation, recombination, and parameter optimization. Each candidate model is evaluated according to two competing objectives: predictive accuracy and model complexity. This results in a Pareto front of solutions that balance goodness of fit, typically quantified by a loss function $L(f)$, and expression complexity $C(f)$, which penalizes the number of operations or functional depth of the model ($f$). The selection process therefore does not yield a single optimal equation, but rather a set of candidate models that achieve different trade-offs between accuracy and simplicity.[35]

Importantly, this procedure is purely data-driven and does not incorporate physical constraints or prior knowledge unless explicitly enforced. Consequently, multiple expressions with comparable loss can differ significantly in their functional form and physical plausibility. In classical workflows, model selection is typically performed by choosing a point along the Pareto front, often based on heuristic criteria or minimal complexity for a given error threshold. However, such selection does not guarantee consistency with underlying physical principles, as the regression process optimizes only statistical fit. This limitation motivates the need for an additional evaluation layer that can assess candidate models in terms of physical validity and interpretability, enabling a transition from purely data-driven fitting to hypothesis-driven reasoning. Complementary to this perspective, recent approaches enhance symbolic regression by incorporating physics-informed constraints[36] or by leveraging large language models to guide equation generation through context-aware reasoning,[37] thereby biasing hypothesis generation itself toward physically plausible models. In contrast, the approach considered here preserves hypothesis generation as a purely data-driven process, while introducing physics at a subsequent evaluation stage to assess and select among candidate models.

### II.2 LLM-Based Evaluation of Physical Consistency

To augment purely data-driven hypothesis generation, we introduce an LLM-based evaluation module that provides an additional mechanism for assessing candidate models beyond statistical fit. Large language models can be viewed as compact representations of a broad corpus of scientific knowledge, acquired through pretraining on diverse textual sources, including scientific literature. As a result, they exhibit strong capabilities in reasoning, abstraction, and synthesis, which can be leveraged to evaluate candidate analytical expressions in the context of established physical principles.

In this work, the LLM is employed as a structured evaluator rather than a generator of hypotheses. Given a set of candidate equations produced by symbolic regression, along with definitions of variables and relevant experimental context, the LLM is tasked with assessing the physical consistency of each model. This includes evaluation of expected trends, limiting behavior, scaling relationships, and qualitative agreement with known physical laws. The output of this evaluation consists of (i) a quantitative score reflecting the degree of physical plausibility and (ii) a concise reasoning that explains the assessment. The scoring provides a mechanism for ranking candidate models, while the reasoning introduces interpretability into the selection process.

The LLM can be deployed in different configurations depending on the application. In the simplest case, a general-purpose pretrained model is used directly. Alternatively, domain adaptation can be introduced through additional conditioning on relevant literature or problem-specific context, enabling more focused evaluation while retaining the ability to generalize beyond the provided information. Importantly, the model is not restricted to a fixed hypothesis space and can incorporate broader physical knowledge when assessing candidate expressions.

A key aspect of this approach is that the generated reasoning is treated as an explicit output of the system rather than an auxiliary byproduct. This reasoning can be propagated within a hierarchical experimental framework and used by downstream decision-making modules or agents to guide further actions. In this way, the LLM-based evaluation not only enables physics-informed ranking of candidate hypotheses but also facilitates knowledge transfer across different levels of an autonomous experimental workflow.

### II.3 Hypothesis Selection and Experimental Decision-Making

The final hypothesis selection is based on a combination of multiple criteria arising from both data-driven and physics-informed evaluations. Specifically, symbolic regression provides predictive loss and model complexity, while the LLM-based module introduces an additional score ($S_{ph}$) reflecting physical consistency. These criteria can be combined in different ways to prioritize candidate models. A simple approach is to define a composite score as a weighted combination of loss, complexity, and LLM-based evaluation: $S_{sum} = \omega_L L(f) + \omega_C C(f) + \omega_{ph} S_{ph}(f)$, where $\omega_L$, $\omega_C$, $\omega_{ph}$ are weights. However, more advanced strategies may be employed depending on the application, including multi-objective ranking or hierarchical selection schemes. Importantly, this flexibility allows the framework to adapt to different regimes where statistical accuracy or physical plausibility may dominate.

Important to notice, that while symbolic regression typically returns models along the Pareto front of loss and complexity, in some cases restricting selection to this set may not be sufficient. Physically consistent models may lie outside the Pareto-optimal set, especially in the presence of noise or limited data. Therefore, more comprehensive implementations may consider an extended candidate space, including suboptimal expressions in terms of loss but potentially more consistent with underlying physics. This highlights that, for real world applicatons, physical consistency may outweigh purely data-driven optimality. We believe that in the short term this approach can be incorporated by adding the different noise realization to experimental data and ensembling the symbolic regression, but access to expressions outside of Pareto will be a preferred strategy.

Once a hypothesis is selected, it is used to guide subsequent experimental decisions. Two model cases can be considered. In the simplest case, the selected model (hypothesis) and its associated reasoning are treated as the final outcome, enabling transition to higher-level decision-making within a broader experimental framework. Alternatively, the process can be iterated to refine the hypothesis through additional data acquisition. In this work, we adopt the latter approach and formulate experiment selection based on residual modeling.

The selected analytical model is used to compute residuals between predicted and experimentally observed values. A Gaussian process (GP) is then trained on these residuals, capturing both their magnitude and associated uncertainty. Regions where the residuals are large indicate discrepancies between the model and experiment, suggesting areas where the current hypothesis is incomplete or inaccurate. To efficiently explore the parameter space and refine the model, BO is applied to the residual model using an acquisition function such as the upper confidence bound (UCB), with an increased exploration component to ensure coverage of the parameter space while retaining sufficient exploitation to refine regions of known discrepancy. This enables targeted acquisition of new data points that are most informative for improving the hypothesis.

This integration of hypothesis selection with residual-driven experiment design establishes a closed-loop process in which models are iteratively refined based on both data consistency and physical plausibility, enabling a transition from static model fitting to adaptive, hypothesis-driven experimentation.

## III. Experimental System and Implementation

### III.1 Ferroelectric Domain Switching as a Model System

Experiments were performed on a lead zirconate titanate (PZT) thin film using scanning probe microscopy (SPM). Ferroelectric domain switching provides a well-established model system for studying nonlinear, field-driven processes, with multiple competing analytical descriptions proposed in the literature. This makes it a suitable testbed for evaluating whether the proposed framework can identify physically consistent and interpretable domain growth laws directly from experimental data.

Local domain switching was induced by applying voltage pulses through a conductive probe at a fixed spatial location (Figure 2). The pulse duration was varied in the range of 0.01–10 s, while the applied voltage ranged from 1 to 10 V. The objective of the experiment was to identify an interpretable analytical relationship between switching conditions and the resulting domain size. Domain structures were visualized using dual AC resonance tracking piezoresponse force microscopy (DART-PFM). Following each switching event, the resulting domain was imaged, and its size was extracted using automated phase analysis. The domain size was quantified through an effective radius defined as $r = \sqrt{A/\pi}$, where $A$ is the measured domain area.

All measurements were conducted in a fully automated regime, with microscope control implemented via the AESPM Python library.[38] The experiment was structured as an iterative loop in which measurement conditions were adaptively updated based on the evolving hypothesis.

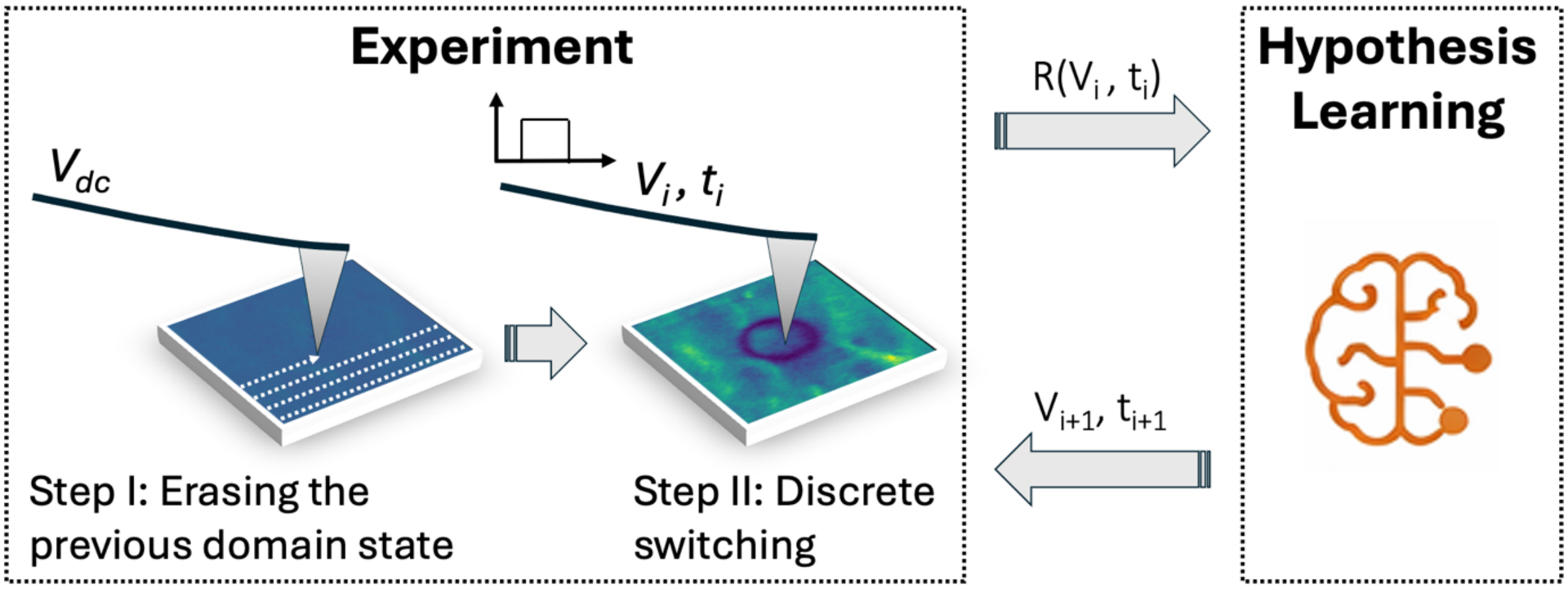

**Figure 2.** Autonomous hypothesis-learning workflow for local ferroelectric domain switching. $V_i$ and $t_i$ denote the switching-pulse parameters: voltage amplitude and pulse duration, respectively.

### III.2 Hypothesis Learning Setup

Symbolic regression was performed using PySR Python library to generate candidate analytical models describing domain growth. The search space was defined using a restricted set of operators, including basic arithmetic operations (+, ×, ÷) and selected nonlinear functions (square, logarithm, exponential, and absolute value). This choice provides mild physical priors while preserving sufficient flexibility to capture diverse functional forms. In this work, candidate models were selected from the Pareto front balancing predictive loss and model complexity.

Candidate expressions were evaluated using Gemini, accessed via API, as a physics-aware evaluation module. Prior to evaluation, the model was provided with a compact summary of established ferroelectric domain-growth models to supply problem-specific physical context. This summary was generated automatically using a GPT-based deep-research pipeline, without manual curation, and supplied to the evaluator as-is (see Data Availability). For each candidate hypothesis, the LLM returned a physical-consistency score together with a brief reasoning statement. The LLM score was then used as the criterion for model selection within the hypothesis-learning loop.

### III.3 Adaptive Experiment Design

The experiment was initialized with 5 randomly selected seed conditions in the (t, V) parameter space. Following initial data acquisition, the system proceeded through 10 iterative cycles. At each iteration, 5 new experimental conditions were selected based on a Gaussian process (GP) model trained on the residuals between experimental observations and predictions of the currently selected hypothesis.

Regions where residuals are large indicate discrepancies between the inferred model and experimental data, highlighting areas where the current hypothesis is incomplete. To efficiently explore these regions, BO was applied to the residual model using UCB acquisition function. This residual-driven adaptive sampling enables iterative refinement of the inferred domain growth model, establishing a closed-loop process in which hypotheses are continuously updated based on both data consistency and physical plausibility.

## IV. Results and Discussion

The practical realization of the initialization step is shown in Figure 3, where the first five domain-switching experiments were performed using randomly selected pulse parameters in the voltage-duration space. These measurements already capture the essential contrast between sub-threshold and above-threshold switching regimes. In the low-voltage, short-duration regime, the applied field is insufficient to nucleate a stable ferroelectric domain, and the extracted effective radius is therefore assigned as zero. At higher voltages and longer pulse durations, switched domains are clearly resolved in the PFM phase response and quantified through automated segmentation. Although five measurements are clearly insufficient to establish a robust physical law for domain growth in the two-dimensional $(V, t)$ space, they provide the minimal experimental basis required to initiate the hypothesis-learning loop. The resulting dataset is used to generate the first set of symbolic-regression models, which then serve as the starting point for BO-driven hypothesis refinement.

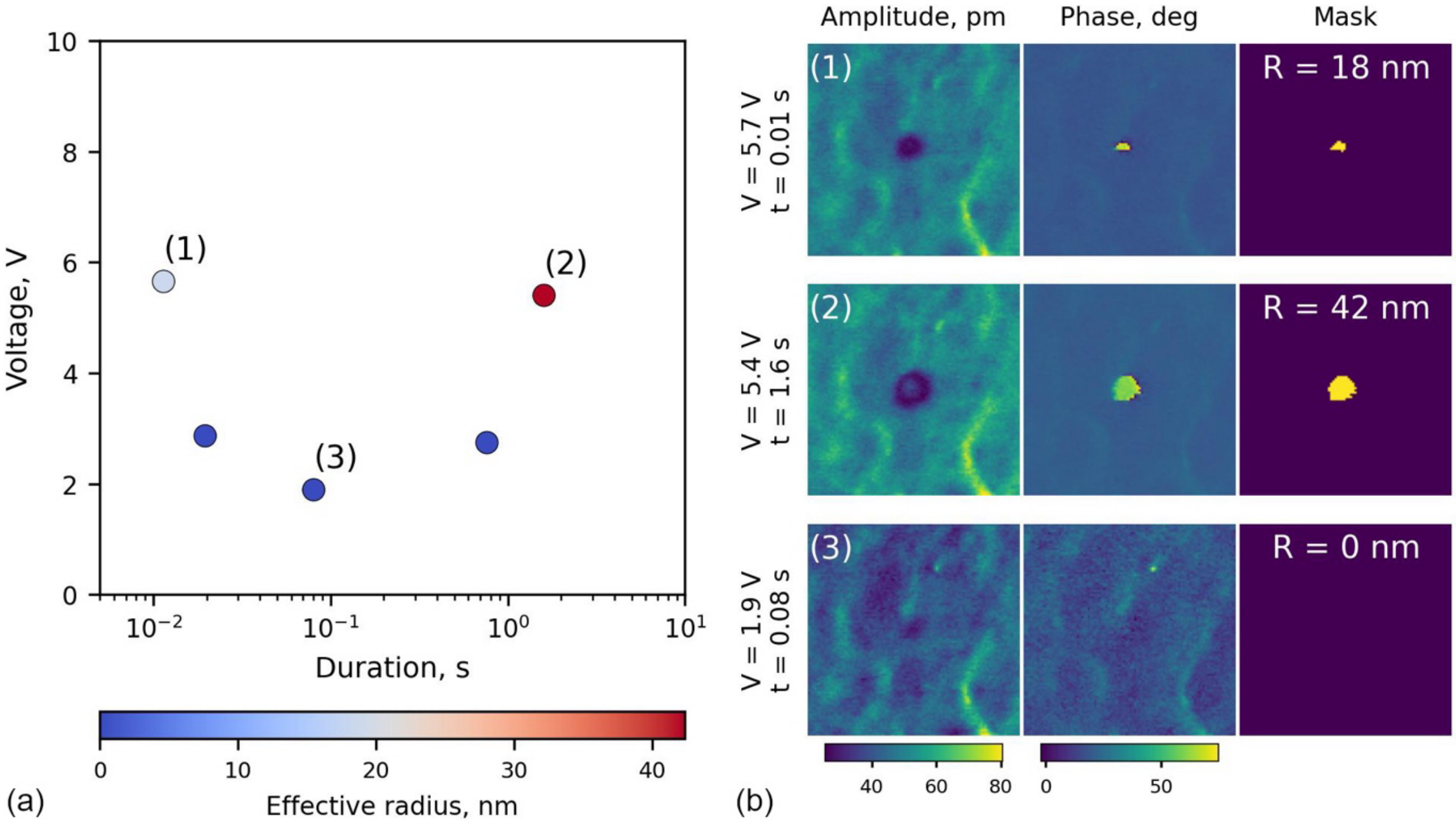

**Figure 3.** Initial seed measurements and domain-radius extraction. (a) Randomly selected seed points in the voltage–pulse-duration space, colored by the extracted effective domain radius. (b) Representative PFM amplitude and phase images with corresponding binary masks used to identify the switched domains and extract the effective radius. The scan size is 700 × 700 nm.

Following random initialization, symbolic regression was trained on the five available measurements to describe the evolution of the effective domain radius in the $(V, t)$ space. The candidate hypotheses generated at this stage, denoted as iteration 0, form a Pareto front in the model loss–complexity space (Figure 4a). This set spans a broad range of expressions, from trivial constant models at the low-complexity end to higher-complexity models with reduced loss that include dependencies on both pulse duration and voltage, as well as threshold-like behavior introduced through the ReLU operator. However, because the dataset is extremely sparse, many of the proposed expressions are not physically meaningful. For example, several models depend only on one variable and ignore the other, such as $r = 0.0216t$ or $r = ReLU(t - 1.5601)$. This is inconsistent with the expected physics of ferroelectric domain growth, where the switched radius should depend on both the applied voltage and pulse duration. These early hypotheses therefore primarily reflect the numerical trade-off between regression loss and expression complexity, rather than a robust physical law. At the same time, they provide the first structured hypothesis space from which the BO-driven refinement cycle can proceed.

After completion of the adaptive measurement cycles, symbolic regression was performed using the full set of 55 measured switching conditions. The resulting Pareto front contains a broader and more physically informative set of candidate expressions than in the initialization step, also ranging from constant and single variable (voltage-only in this case) models to coupled voltage–time growth laws (Figure 4b). The lowest-complexity expression, $r = 0.035$, represents only the mean response and has no physical content as a switching law. The voltage-only models, $r = 0.0067V$ and $r = 0.0067V + 0.00024$, capture the expected increase of domain size with applied bias, but they neglect pulse duration and therefore can only be interpreted as highly reduced equilibrium or bias-controlled approximations. This is incomplete for finite-time domain writing, where both nucleation and domain-wall motion contribute to the observed radius. More physically plausible hypotheses appear at higher complexity, where symbolic regression identifies coupled forms such as $r = (0.00052t + 0.0058)V$, $r = (0.00194\sqrt{t} + 0.00493)V$, and $r = (0.000786\log t + 0.00781)V$. The linear-in-time form is monotonic but likely over-simplified, since standard models of tip-induced switching rarely predict unrestricted linear radial growth. In contrast, the $\sqrt{t}$-dependent expression resembles transport- or screening-limited growth,[39] where the time dependence can approach $b \approx 1/2$ in simplified diffusion-limited limits.[40] In our case, the nearly linear dependence on voltage suggests an empirical growth law of the form $r \sim (V - V_0)t^b$, consistent with prior observations of approximately linear voltage scaling in AFM-induced domain growth, while the sublinear time exponent reflects the slowing of growth as screening, conductivity, and the local driving field evolve.[31, 41] The logarithmic expression is also physically meaningful. In the creep regime, the domain wall velocity depends exponentially on the inverse local electric field. Because the tip-induced field decreases with increasing domain size, the growth rate becomes progressively suppressed, leading to an effective logarithmic-in-time domain expansion.[31, 42] The experimentally measured domain radii are compared with the values predicted by candidate models from the first and final iterations in Figure 4e.

Selection of a single expression from the symbolic-regression Pareto front was first performed using the internal PySR "best" criterion (Figure 4c,d). Each candidate expression is assigned a score that quantifies the reduction in logarithmic loss per added unit of expression complexity, $S = -d\log(\mathcal{L})/dC$, where $\mathcal{L}$ is the regression loss and $C$ is the expression complexity. The best criterion then selects the highest-scoring expression among models whose loss remains close to the minimum-loss solution. Therefore, this selection rule provides a quantitative trade-off between accuracy and parsimony, but it does not explicitly evaluate physical validity. At iteration 0, the PySR-selected model was $r = \mathrm{ReLU}[0.0175(t + V - 4.5927)]$ (Figure 4c). Among the early candidate expressions, this model is relatively physically grounded because it depends on both pulse duration and applied voltage and introduces an effective threshold through the ReLU function. In contrast, after the final iteration, the PySR best criterion selected the voltage-only expression $r = 0.00671V$ (Figure 4d). This occurs because the transition from a constant model to a simple voltage-dependent model produces a large reduction in loss with only a small increase in complexity, whereas moving further along the Pareto front toward coupled $V$–$t$ expressions increases complexity while producing only a comparatively modest additional reduction in loss. As a result, the purely quantitative PySR selection favors a parsimonious voltage-only law, even though it neglects pulse duration and is incomplete as a finite-time domain-growth model. This contrast highlights the need for an additional physics-based evaluator capable of distinguishing statistically efficient expressions from physically meaningful hypotheses.

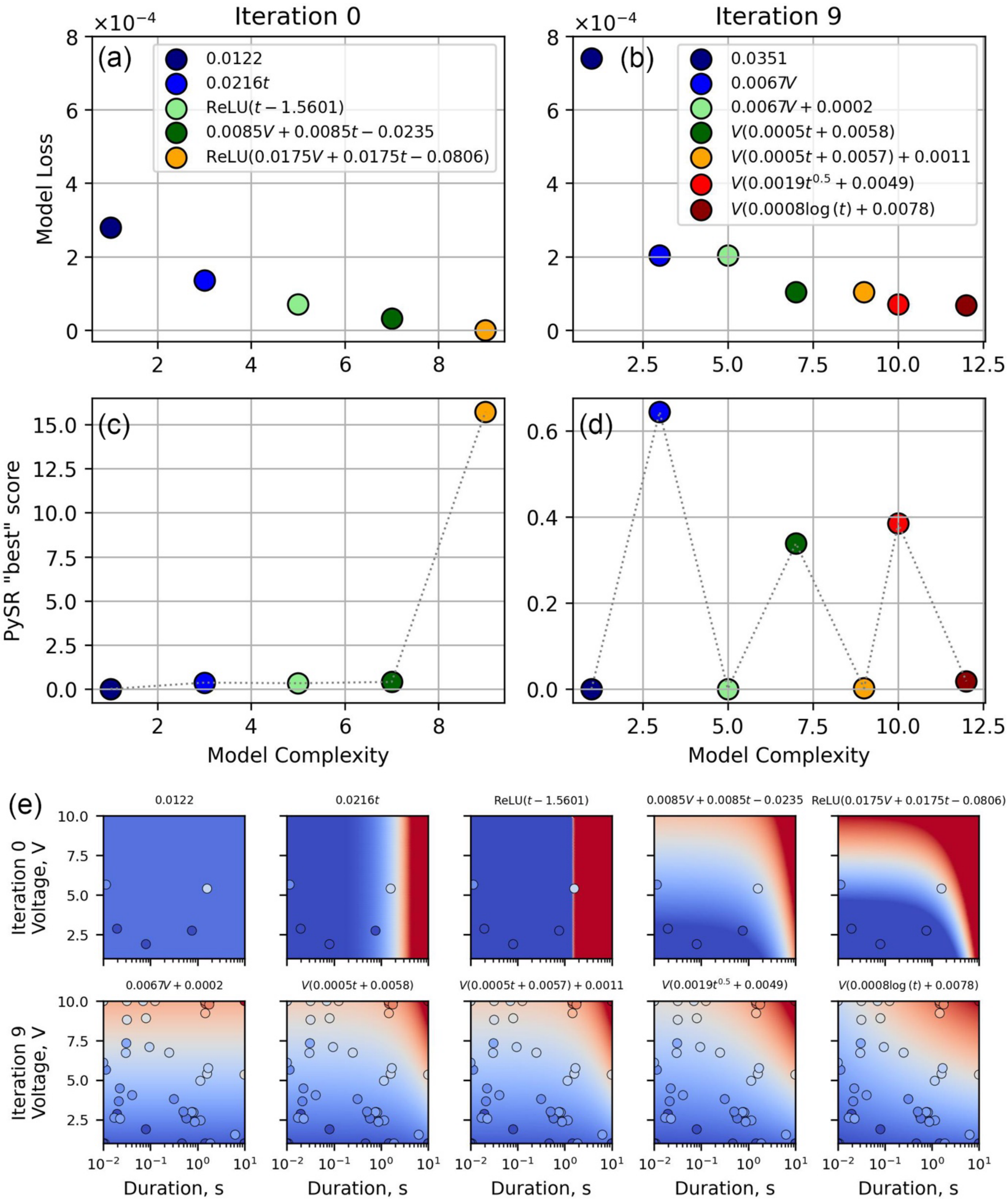

**Figure 4.** (a,b) Pareto front in the loss-complexity space of symbolic regression models at early (iteration 0) and late (iteration 9) stages. (c,d) Corresponding PySR "best" scores, illustrating purely data-driven prioritization of candidate expressions. (e) Predicted effective-radius maps in the voltage–pulse-duration space for the candidate equations at iterations 0 and 9, with experimentally measured points overlaid. As the dataset grows, the hypothesis space evolves from simple or physically incomplete expressions toward coupled voltage–time models with more interpretable domain-growth behavior.

A physics-aware LLM evaluator can be constructed as a structured scoring module that receives candidate symbolic expressions together with variable definitions, experimental context, and a compact representation of relevant physical knowledge:

```
You are a Physics Validator for Ferroelectric Domain Growth.
Inputs: x0 = pulse duration t, x1 = voltage V. Target: domain radius r.

SOURCE USAGE:
Use the provided document as structured guidance on the main classes of physically relevant models for ferroelectric domain growth under
an SPM tip. In particular, the document organizes domain growth laws into three key families:
(i) thermodynamic / bias-controlled equilibrium scaling,
(ii) kinetic growth via domain-wall motion in a decaying tip field,
(iii) disorder- or screening-controlled regimes (e.g. logarithmic or creep behavior).

Use these model families as reference templates for judging whether a candidate equation is physically plausible. However, do NOT require
an equation to exactly match any specific formula in the document. The document is not exhaustive, and realistic systems often involve
combinations of mechanisms rather than a single closed-form law.

Evaluate each candidate primarily using general physical principles (electrostatics, activated kinetics, thermodynamics, and screening),
while using the document to:
- check consistency of scaling trends (e.g., r ~ V^(1/3), slow/logarithmic time growth),
- recognize known mechanisms (Merz-like activation, LGD scaling, creep behavior),
- identify whether the equation plausibly corresponds to one or a combination of recognized physical regimes.

An equation may receive a high score even if it is not explicitly present in the document, provided it is consistent with these physical
mechanisms. Conversely, equations that fit numerically but do not correspond to any plausible physical regime should be penalized.

SCORING CRITERIA:
10 = physically consistent, interpretable, correct scaling
7-9 = mostly correct, minor issues
4-6 = partially correct but flawed
1-3 = weak or inconsistent
0 = unphysical or trivial

PHYSICS RULES:

1. NON-TRIVIALITY:
- Must depend on BOTH t and V.
- Constant or single-variable dependence → score ≤ 2.

2. MONOTONICITY:
- r should increase with V and t.
- Violation → strong penalty.

3. FIELD SCALING:
- Electric field decays as E ~ r^-3.
- Reject equations implying increasing field with r.

4. REGIME BEHAVIOR:
- Early-time: activated kinetics (e.g. exp(-alpha/E))
- Late-time: slower saturation / thermodynamic scaling
- Reward equations capturing transition between regimes.

5. DIVERGENCE / PATHOLOGY:
- Penalize singularities, exploding exponentials, log(log(...)), etc.

6. INTERPRETABILITY:
- Prefer simple, physically interpretable expressions.
- Penalize unnecessary complexity.

FINAL SCORE:
score = physics_score - complexity_penalty
Clamp score to [0,10].
```

Closely related approaches have recently integrated LLMs directly into symbolic-regression workflows by adding an LLM-derived term to the regression loss function, where the model scores candidate equations for dimensional consistency, simplicity, and physical realism; in that case, the LLM acts as a regularizer that biases the search process itself.[36] In contrast, we keep symbolic regression as a data-driven hypothesis generator and use the LLM only after candidate equations are produced, as an independent evaluator of physical plausibility. The evaluator was initialized with a GPT-generated deep-research summary of ferroelectric domain-growth models, which served as a lightweight domain-conditioning step and emulated refinement of the evaluator agent toward the specific physical problem (see Data Availability). This conditioning is not equivalent to fine-tuning or expert curation, and it could be improved by using a verified literature database, human-reviewed physical rules, multiple evaluator agents, or retrieval from primary sources. The prompt then defines the variables $(x_0 = t, x_1 = V)$, target quantity $r$, relevant model families, and a scoring rubric based on non-triviality, monotonicity, field scaling, limiting-regime behavior, pathological divergence, and interpretability. This structure provides a transparent place to inject human physical knowledge into the autonomous loop. For example, the rule that a finite-time growth law should depend on both $t$ and $V$encodes a reasonable expectation for domain switching, but it may also overconstrain the evaluator by penalizing valid limiting cases such as equilibrium voltage-controlled scaling. Therefore, the score should be interpreted as a heuristic measure of physical plausibility rather than proof of correctness. Importantly, the evaluator returns both a quantitative score and a short reasoning trace. The score enables automated ranking and combination with regression metrics, while the reasoning makes the selection interpretable, exposes the physical assumptions behind the ranking, and allows the output to be propagated to downstream decision-making agents. This separation between data-driven hypothesis generation and physics-aware evaluation provides a flexible framework: future implementations can refine the prompt, calibrate scores against expert annotations, use ensembles of LLMs, or replace heuristic rules with formal checks such as dimensional analysis, monotonicity tests, and uncertainty-aware verification.

In the present implementation, the adaptive experiment was driven by the hypothesis selected solely according to the LLM-based physical-plausibility score at each iteration, while symbolic regression serves as a method for the hypotheses generation. The first candidate equations generated from the five seed measurements received generally low LLM scores, consistent with their limited physical content and frequent violation of basic expectations for finite-time domain growth (Figure 5a). In particular, several early expressions were constant or single-variable and therefore failed the basic criteria encoded in the evaluator prompt. By contrast, the candidate equations obtained after the final iteration received substantially higher LLM scores, reflecting the emergence of models that depend on both voltage and pulse duration and preserve physically expected monotonic trends (Figure 5b). Based on the evaluator output, satisfying these basic physical constraints provides the foundation for an intermediate-to-high score, while recognizable similarity to established limiting laws, such as screening-limited or creep-like domain growth, further increases the ranking. The LLM evaluator assigned the highest score to $r = (0.000786 \log t + 0.00781)V$, because this expression combines voltage-assisted growth with logarithmic time dependence, consistent with slow domain-wall motion in the creep regime under the decaying electric field of an SPM tip. Thus, the LLM-based selection promotes hypotheses that are not only statistically plausible but also interpretable within known physical mechanisms of ferroelectric domain growth.

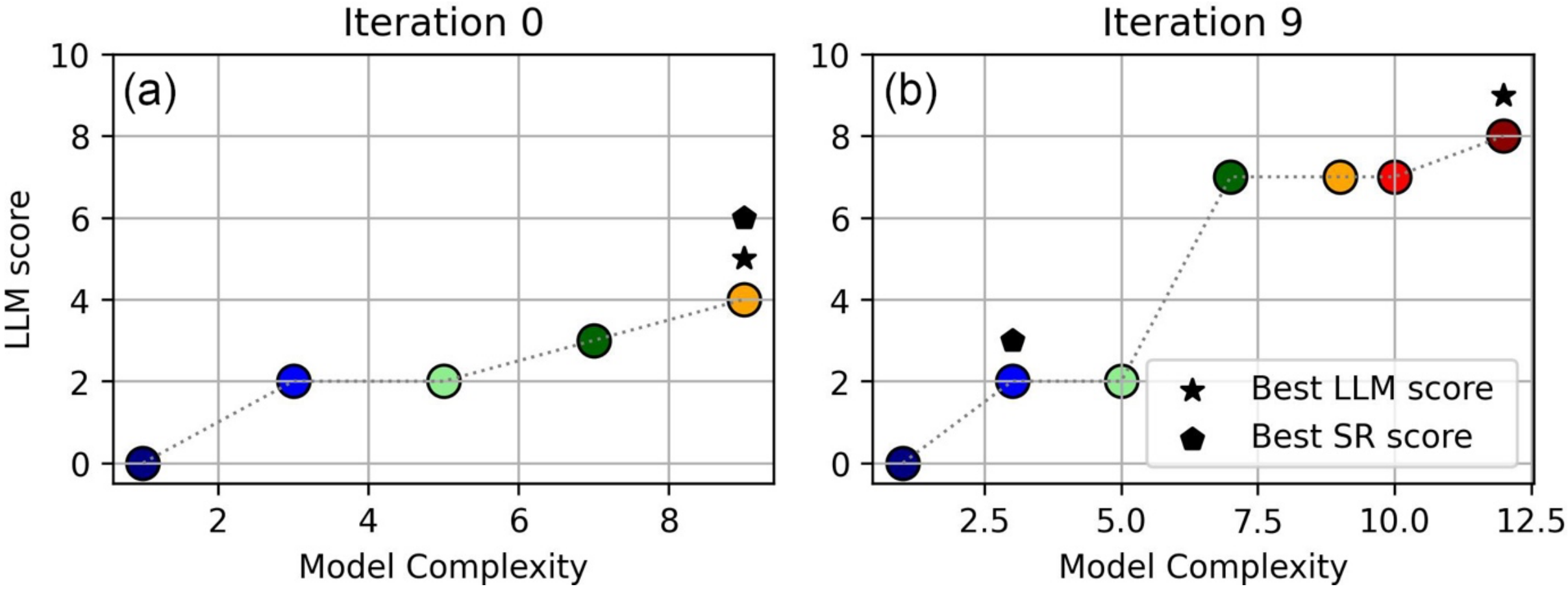

**Figure 5.** LLM-based scoring of symbolic regression models. LLM scores of candidate models at early (iteration 0) and late (iteration 9) stages, shown as a function of model complexity. Stars denote the model selected by the LLM-based score, while pentagons indicate the model preferred by symbolic regression.

The distinction between statistical fitting and physically meaningful hypothesis discovery is most clearly reflected in the ability of a selected model to generalize beyond the data used for its selection. In the present experiment, we do not evaluate extrapolation outside the measured voltage–duration range. Instead, we use a forward-looking validation scheme: at each iteration, models selected by three different criteria – minimum symbolic-regression loss, PySR best score, and LLM-based physical-plausibility score – are evaluated both on the data available at the time of selection and on the subsequently acquired "future" points that were not used for training at that iteration (Figure 6).

On the available training data (Figure 6a,c), the model selected by the minimum SR loss gives the lowest MSE at each iteration, as expected, since this criterion explicitly optimizes agreement with the measured points. In contrast, the model selected by the PySR best score shows the largest discrepancy with the experimental data. This behavior reflects the role of the complexity penalty in the PySR selection rule. PySR does not select models solely by minimizing prediction error; rather, it favors equations that achieve a favorable balance between accuracy and simplicity. Therefore, a model with somewhat larger absolute error can still be selected if it is substantially simpler. In contrast, the LLM-selected model shows intermediate training error, indicating that physics-guided selection is not equivalent to choosing the best numerical fit. Instead, it prioritizes expressions that are consistent with basic physical constraints of the switching process while remaining interpretable.

Evaluation on the future points provides a more relevant test of hypothesis quality (Figure 6b,d), since these measurements were not available at the time of model selection. In this case, the difference between the MSE of the LLM-selected models and the minimum-loss SR models is substantially smaller than on the training data. This indicates that physics-based scoring can identify models that remain predictive under subsequent measurements, even when they are not the best interpolators of the currently available dataset. In several iterations, the LLM-selected hypothesis clearly outperforms the loss-selected model on future points, while in the remaining iterations it shows comparable performance. When averaged across all iterations, the loss-based and LLM-based selections therefore exhibit similar predictive accuracy, with the LLM-selected models showing lower variation in performance (Figure 6d). We note that this comparison should not be interpreted as a rigorous statistical validation: the future points were not sampled randomly, but were selected by BO on the residuals, and the number of remaining future points decreases with iteration number. Nevertheless, at a qualitative level, this analysis shows that the physics-based evaluator is consistent with subsequent experimental observations and can select physically meaningful hypotheses without sacrificing predictive performance within the explored parameter space.

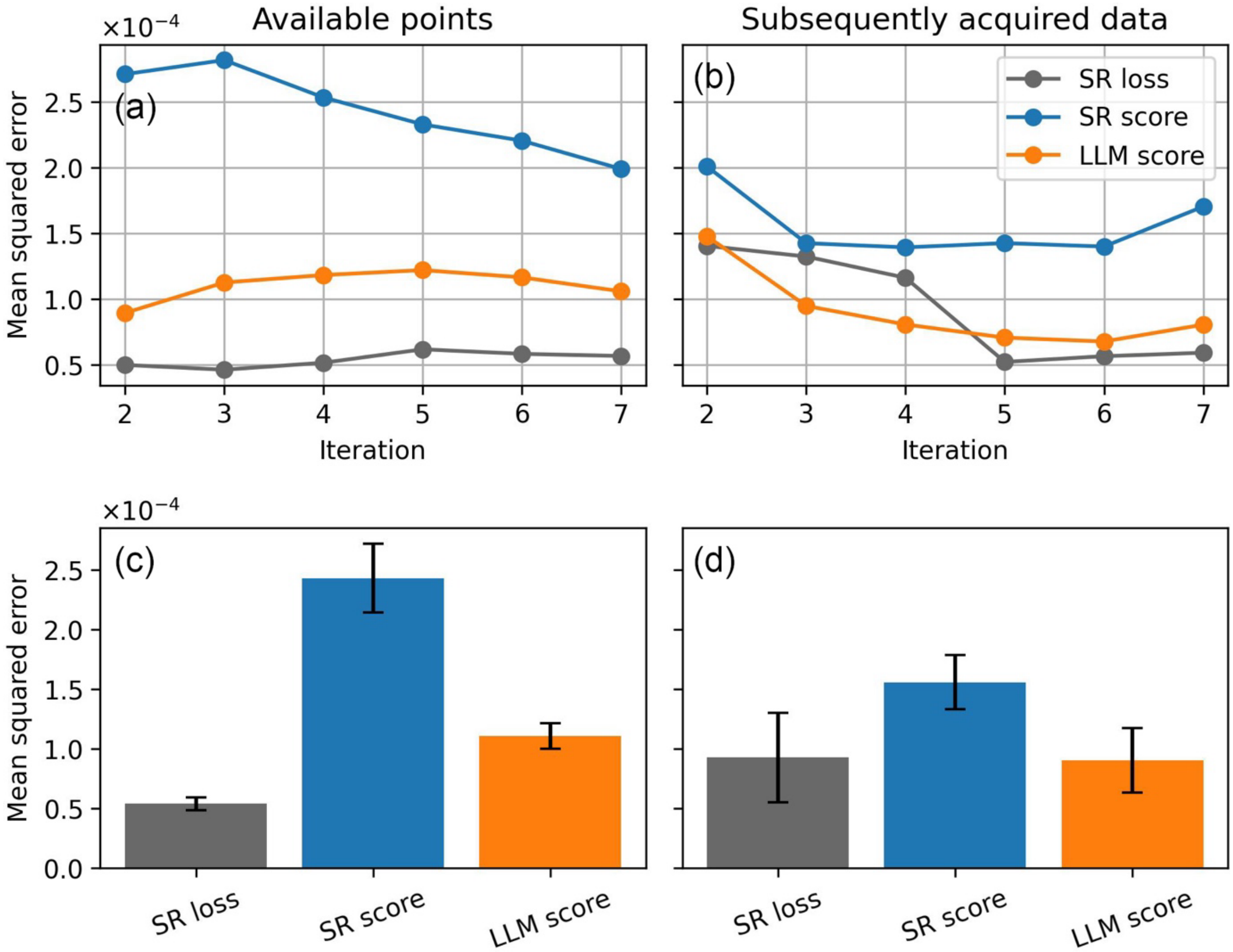

**Figure 6.** Mean squared error between model predictions and measured values for models selected using different criteria. (a,b) MSE evaluated on data available at the time of selection and on subsequently acquired measurements, respectively, shown as a function of iteration. (c,d) Corresponding average MSE across iterations for available and subsequently acquired data.

The reasoning component is as important as the quantitative LLM score because it converts model selection from a black-box ranking into an interpretable hypothesis-assessment step. This is illustrated in **Table 1**, where the voltage-only model, $r = 0.0067V$, receives a low score despite its favorable PySR score, because the evaluator identifies the absence of pulse-duration dependence as a critical limitation for a finite-time domain-growth law. The LLM evaluation results for all equations inferred by symbolic regression during the automated experiment are provided in the Supplementary Materials. The reasoning does not classify the expression as entirely meaningless: it recognizes that approximately linear voltage scaling may be compatible with simplified bias-controlled or near-threshold regimes, but concludes that the model is incomplete as a kinetic growth hypothesis. In contrast, for the model $r = V(0.0008logt + 0.0078)$, the reasoning explains the high score by explicitly linking the expression to kinetic growth in a disorder-controlled or creep-like regime.

This natural-language reasoning is particularly important for hierarchical autonomous experimentation. In such workflows, the outcome of one loop is not only a selected action or numerical objective, but also a piece of interpretable knowledge that must be propagated across different stages, decision layers, and agents. A scalar score can rank hypotheses, but it does not explain why a model should be trusted, what regime it represents, or where its limitations are. In contrast, the reasoning text provides a common representation that can be used by downstream modules: it identifies the relevant physical mechanism, flags violated assumptions, distinguishes incomplete limiting cases from unphysical models, and provides a basis for subsequent experimental planning. Thus, the LLM evaluator serves not only as a scoring function but also as a knowledge-transfer interface between symbolic regression, BO, and higher-level agentic control.

**Table 1.** LLM-based reasoning for candidate models. Comparison of low- and high-scoring models with corresponding LLM evaluation.

| | $\boldsymbol{R = 0.0067V}$ | $\boldsymbol{R = V(0008 * \log(t) \ + 0.0078)}$ |
|---|---|---|
| **Score** | **2/10** | **8/10** |
| **Matching model** | Thermodynamic / Bias-Controlled Equilibrium Scaling (simplified) | Kinetic Growth (Disorder-controlled Creep) |
| **Reasoning** | `This equation depends only on voltage (x1) and not on pulse duration (x0), violating Physics Rule 1 (Non-triviality) for a domain *growth* model. While a linear dependence of radius on voltage (r ~ V) can be observed in some kinetic-limited or near-threshold regimes (as mentioned on page 10 for 'r ~ V' regimes), the complete absence of time dependence makes it an incomplete model for domain growth kinetics. It is plausible only as an instantaneous equilibrium size, not a growth law, and thus receives a low score.` | `This equation can be written as r = (C1 * log(t) + C2) * V, where C1 = 0.0007866032 and C2 = 0.0078091975 are positive constants. It depends on both t and V. The logarithmic time dependence (`r ~ log(t)`) is a key signature of 'disorder-controlled kinetics' or 'creep behavior' when combined with a decaying tip field, as explicitly mentioned in the document (page 1 and page 5: `r(t) ~ V^(1/n) * [ln(t/t0)]^(1/(ημ))`). The `r ~ V` scaling is common. Monotonicity is generally maintained for `t > exp(-C2/C1)` (approximately 0.00005, which is typically outside the relevant physical range for creep). This is a strong and physically interpretable model, closely matching a recognized regime, and thus receives a high score.` |

A direct comparison with the previous hypothesis-learning SPM study[31] shows strong physical consistency between the two results, despite the different hypothesis-generation strategies. In that work, the autonomous experiment on an 80 nm $BaTiO_3$ film selected the kinetic domain-wall-motion model, $r(V,t) = V^a \log\ t$, as the dominant hypothesis; the authors interpreted this as evidence that domain growth was controlled primarily by domain-wall pinning/creep rather than by thermodynamic equilibrium size or screening-charge transport.[31] In the present work on a PZT film, the LLM-based evaluator selected $r = V(0.0008\log\ t + 0.0078)$, which has the same essential structure: voltage-assisted growth with logarithmic time dependence. The correspondence is not exact – the earlier model includes a fitted voltage exponent $a$, whereas our symbolic-regression result is approximately linear in $V$ with an additive offset in the logarithmic prefactor – but both expressions point to the same physical regime: slow kinetic growth of a ferroelectric domain wall in the strongly inhomogeneous field of an SPM tip. This agreement is important because in the earlier work the logarithmic kinetic model was predefined by experts as one of the competing hypotheses, while in our case a closely related expression emerged from the data-driven symbolic-regression search and was selected afterward by the LLM evaluator. Thus, the comparison supports the central physical conclusion that the observed switching is governed by kinetic, creep-like domain-wall motion, while also demonstrating the methodological advance from selecting among predefined hypotheses to generating and evaluating candidate laws directly from experimental data.

A broader opportunity opened by this workflow is the tighter integration of symbolic hypothesis generation with optimization and model calibration. In the present implementation, symbolic regression provides analytical candidate forms, while Bayesian optimization is used to select new experiments based on the residuals of the currently selected hypothesis. Future implementations could close this loop more directly by using symbolic regression to propose compact model structures and then optimizing their parameters, validity ranges, and uncertainty estimates within the autonomous experiment. At the same time, not all physically meaningful hypotheses admit simple symbolic representations. Many relevant models in microscopy and materials science are numerical, simulation-based, or depend on internal state variables that cannot be reduced to a closed-form expression. For such cases, the same hypothesis-learning logic can be extended beyond symbolic equations. Expensive models can be incorporated through co-navigation strategies, where experiments and computational models are jointly updated through surrogate models or reduced-dimensional parameter spaces.[43, 44] Cheaper non-symbolic models can be calibrated directly by BO or Bayesian inference over a low-dimensional set of physical parameters, and multiple model classes can be treated as competing hypotheses in the loop. In this view, symbolic regression represents one practical and interpretable route to open hypothesis generation, but the broader framework is more general: autonomous experimentation can generate, compare, calibrate, and refine both analytical and computational hypotheses as part of a unified discovery process.

## Conclusion

We introduce an open hypothesis-learning workflow for autonomous SPM experiments, in which symbolic regression generates candidate analytical laws directly from data and an LLM-based evaluator ranks them according to physical plausibility. The approach was implemented for ferroelectric domain switching in a PZT thin film. Starting from five randomly selected seed measurements in voltage–pulse-duration space, symbolic regression produced an initial Pareto front of candidate equations. As residual-driven Bayesian optimization acquired additional data, the hypothesis space evolved from mostly trivial or incomplete expressions toward physically meaningful voltage–time growth laws. The active hypothesis-learning process identified kinetic, creep-like domain-wall growth in the decaying electric field of an SPM tip, consistent with prior hypothesis-learning SPM work where a predefined $r(V,t) = V^{a}\log t$ model was selected for domain growth. Importantly, in the present workflow this model class emerged from data-driven symbolic regression rather than being specified in advance. Overall, this work positions symbolic regression combined with LLM-based reasoning as a route from closed-loop optimization toward open hypothesis discovery. Future extensions can incorporate expert-calibrated evaluators, formal physical checks, numerical models, and multiple hypothesis classes within the same autonomous loop.

**Acknowledgments**
This material is based upon work supported by the National Science Foundation under Award No. NSF 2523284.

**Author contributions**
BNS: Conceptualization; Software; Data curation; Writing – original draft. UP: Writing – review & editing. YL: Software; Writing – review & editing. KB: Writing – review & editing. SVK: Conceptualization; Supervision; Writing – review & editing.

**Data Availability Statement**

The analysis codes and results that support the findings of this study are available at https://github.com/Slautin/2026_domain_growth_hypothesis_learning.

1. Tom, G. et al. Self-driving laboratories for chemistry and materials science. Chem. Rev. 124, 9633–9732 (2024).

2. Stach, E. et al. Autonomous experimentation systems for materials development: a community perspective. Matter 4, 2702–2726 (2021).

3. Spurgeon, S. R. et al. Towards data-driven next-generation transmission electron microscopy. Nat. Mater. 20, 274–279 (2021).

4. Roccapriore, K. M., Dyck, O., Oxley, M. P., Ziatdinov, M. & Kalinin, S. Automated experiment in 4D-STEM: exploring emergent physics and structural behaviors. ACS Nano 16, 7605–7614 (2022).

5. Kalinin, S. V. et al. Automated and autonomous experiments in electron and scanning probe microscopy. ACS Nano 15, 12604–12627 (2021).

6. Adams, F., McDannald, A., Takeuchi, I. & Kusne, A. G. Human-in-the-loop for Bayesian autonomous materials phase mapping. Matter 7, 697–709 (2024).

7. Rooney, M. B. et al. A self-driving laboratory designed to accelerate the discovery of adhesive materials. Digit. Discov. 1, 382–389 (2022).

8. Lo, S. et al. Review of low-cost self-driving laboratories in chemistry and materials science: the “frugal twin” concept. Digit. Discov. 3, 842–868 (2024).

9. Hysmith, H. et al. The future of self-driving laboratories: from human-in-the-loop interactive AI to gamification. Digit. Discov. 3, 621–636 (2024).

10. Brown, A. K., Soni, A., Lin, D. & Berlinguette, C. P. Accelerated emergence of self-driving laboratories for accelerating materials discovery. ACS Cent. Sci. 12, 300–306 (2026).

11. Ziatdinov, M., Liu, Y., Kelley, K., Vasudevan, R. & Kalinin, S. V. Bayesian active learning for scanning probe microscopy: from Gaussian processes to hypothesis learning. ACS Nano 16, 13492–13512 (2022).

12. Um, M. et al. Tailoring molecular space to navigate phase complexity in Cs-based quasi-2D perovskites via gated-Gaussian-driven high-throughput discovery. Adv. Energy Mater. 15, 2404655 (2025).

13. Liu, Y. et al. Automated materials discovery platform realized: scanning probe microscopy of combinatorial libraries. Preprint at https://doi.org/10.48550/arXiv.2412.18067 (2024).

14. Liang, Q. et al. Benchmarking the performance of Bayesian optimization across multiple experimental materials science domains. npj Comput. Mater. 7, 188 (2021).

15. Kusne, A. G. et al. On-the-fly closed-loop materials discovery via Bayesian active learning. Nat. Commun. 11, 5966 (2020).

16. Gongora, A. E. et al. A Bayesian experimental autonomous researcher for mechanical design. Sci. Adv. 6, eaaz1708 (2020).

17. Foadian, E. et al. ACCEL: automated closed-loop co-optimization and experimentation learning enables phase-pure identification in formamidinium-based Dion–Jacobson halide perovskites. Preprint at https://doi.org/10.26434/chemrxiv-2026-8c93 (2026).

18. Harris, S. B., Vasudevan, R. & Liu, Y. Active oversight and quality control in standard Bayesian optimization for autonomous experiments. npj Comput. Mater. 11, 23 (2025).

19. Jamali, V., Aghazadeh, A. & Kacher, J. Thinking microscopes: agentic AI and the future of electron microscopy. npj Comput. Mater. 12, 149 (2026).

20. Yang, H., Yue, S. & He, Y. Auto-GPT for online decision making: benchmarks and additional opinions. Preprint at https://doi.org/10.48550/arXiv.2306.02224 (2023).

21. Bran, A. M., Cox, S., Schilter, O., Baldassari, C., White, A. D. & Schwaller, P. Augmenting large language models with chemistry tools. Nat. Mach. Intell. 6, 525–535 (2024).

22. Zhou, L. et al. Autonomous agents for scientific discovery: orchestrating scientists, language, code and physics. Preprint at https://doi.org/10.48550/arXiv.2510.09901 (2025).

23. Mitchener, L. et al. Kosmos: an AI scientist for autonomous discovery. Preprint at https://doi.org/10.48550/arXiv.2511.02824 (2025).

24. Wu, Q. et al. AutoGen: enabling next-gen LLM applications via multi-agent conversation. Preprint at https://doi.org/10.48550/arXiv.2308.08155 (2023).

25. Mandal, I. et al. Evaluating large language model agents for automation of atomic force microscopy. Nat. Commun. 16, 9104 (2025).

26. Yamada, Y. et al. The AI Scientist-v2: workshop-level automated scientific discovery via agentic tree search. Preprint at https://doi.org/10.48550/arXiv.2504.08066 (2025).

27. Tang, J., Xia, L., Li, Z. & Huang, C. AI-Researcher: autonomous scientific innovation. Preprint at https://doi.org/10.48550/arXiv.2505.18705 (2025).

28. Zhang, H., Song, Y., Hou, Z., Miret, S. & Liu, B. HoneyComb: a flexible LLM-based agent system for materials science. Preprint at https://doi.org/10.48550/arXiv.2409.00135 (2024).

29. Yao, L. et al. Operationalizing serendipity: multi-agent AI workflows for enhanced materials characterization with theory-in-the-loop. Preprint at https://doi.org/10.48550/arXiv.2508.06569 (2025).

30. Song, T. et al. A multiagent-driven robotic AI chemist enabling autonomous chemical research on demand. J. Am. Chem. Soc. 147, 12534–12545 (2025).

31. Liu, Y. et al. Autonomous scanning probe microscopy with hypothesis learning: exploring the physics of domain switching in ferroelectric materials. Patterns 4, 100704 (2023).

32. Ziatdinov, M. A. et al. Hypothesis learning in automated experiment: application to combinatorial materials libraries. Adv. Mater. 34, 2201345 (2022).

33. Boiko, D. A., MacKnight, R., Kline, B. & Gomes, G. Autonomous chemical research with large language models. Nature 624, 570–578 (2023).

34. Liu, Y., Checa, M. & Vasudevan, R. K. Synergizing human expertise and AI efficiency with language model for microscopy operation and automated experiment design. Mach. Learn.: Sci. Technol. 5, 02LT01 (2024).

35. Cranmer, M. Interpretable machine learning for science with PySR and SymbolicRegression.jl. Preprint at https://doi.org/10.48550/arXiv.2305.01582 (2023).

36. Fox, C., Tran, N. D., Nacion, F. N., Sharlin, S. & Josephson, T. R. Incorporating background knowledge in symbolic regression using a computer algebra system. Mach. Learn.: Sci. Technol. 5, 025057 (2024).

37. Sharlin, S. & Josephson, T. R. In-context learning and reasoning for symbolic regression with large language models. Preprint at https://doi.org/10.48550/arXiv.2410.17448 (2024).

38. Liu, R. aespm: Python interface for automated experiments on scanning probe microscopes. GitHub https://github.com/RichardLiuCoding/aespm (2024).

39. Brugère, A., Gidon, S. & Gautier, B. Finite element method simulation of the domain growth kinetics in single-crystal LiTaO3: role of surface conductivity. J. Appl. Phys. 110, 052016 (2011).

40. Yudin, P. V., Hrebtov, M. Y., Dejneka, A. & McGilly, L. J. Modeling the motion of ferroelectric domain walls with the classical Stefan problem. Phys. Rev. Appl. 13, 014006 (2020).

41. Ievlev, A. V., Morozovska, A. N., Shur, V. Y. & Kalinin, S. V. Humidity effects on tip-induced polarization switching in lithium niobate. Appl. Phys. Lett. 104, 092908 (2014).

42. Paruch, P. & Guyonnet, J. Nanoscale studies of ferroelectric domain walls as pinned elastic interfaces. C. R. Phys. 14, 667–684 (2013).

43. Kennedy, M. C. & O'Hagan, A. Bayesian calibration of computer models. J. R. Stat. Soc. Ser. B Stat. Methodol. 63, 425–464 (2001).

44. Slautin, B. N. et al. Bayesian conavigation: dynamic designing of the material digital twins via active learning. ACS Nano 18, 24898–24908 (2024).